\documentclass[12pt]{article}
\usepackage{graphicx}

\newcommand\pubnumber{SNSN-323-63}
\newcommand\pubdate{\today}

\def\cern{CERN, CH-1211, Geneva-23, Switzerland.}

\def\Title#1{\begin{center} {\Large #1 } \end{center}}
\def\Author#1{\begin{center}{ \sc #1} \end{center}}
\def\Address#1{\begin{center}{ \it #1} \end{center}}

\newcommand\pubblock{\rightline{\begin{tabular}{l} \pubnumber\\
         \pubdate  \end{tabular}}}
\newenvironment{Abstract}{\begin{quotation}  }{\end{quotation}}
\newenvironment{Presented}{\begin{quotation} \begin{center} 
             PRESENTED AT\end{center}\bigskip 
      \begin{center}\begin{large}}{\end{large}\end{center} \end{quotation}}

\def\beq{\begin{equation}}
\def\eeq#1{\label{#1}\end{equation}}
\def\eeqn{\end{equation}}

\def\beqa{\begin{eqnarray}}
\def\eeqa#1{\label{#1}\end{eqnarray}}
\def\eeqan{\end{eqnarray}}

\let\bar=\overbar

\def\Dslash{\not{\hbox{\kern-4pt $D$}}}
\def\dslash{\not{\hbox{\kern-2pt $\del$}}}

\def\msb{{\bar{\ssstyle M \kern -1pt S}}}

\begin{document}
\begin{titlepage}
\pubblock

\vfill
\Title{Reconstruction of primary vertices in $pp$ collisions
at energies of $900$~GeV and $7$~TeV with the ATLAS detector}
\vfill
\Author{ Kirill Prokofiev on behalf of the ATLAS Collaboration}
\Address{\cern}
\vfill
\begin{Abstract}
 The Large Hadron Collider (LHC) of the European Organisation 
 for Nuclear Research (CERN) started its operation in Autumn of 2009.
 The initial run at a centre-of-mass energy of $900$~GeV, has 
 been followed by the on-going run at the energy of $7$~TeV.
 
 While initially the probability of several proton-proton collisions 
 to happen within the same bunch-crossing was
 approximately $10^{-5}$, the level of the pile-up grows steadily
 with better focusing and squeezing of the LHC beams at collision point.
 
 Presented in this contribution is the  performance of the primary 
 vertex reconstruction algorithms used for analysis of the first 
 collisions at the  LHC. Different approaches used for the reconstruction 
 of primary vertices in $900$~GeV and $7$~TeV collisions are presented.
 The efficiencies of the primary vertex reconstruction used for the 
 first physics analyses of ATLAS are shown.  The resolutions on 
 positions of the reconstructed primary vertices are investigated
 by studying the distributions of pulls of distance between 
 artificially created half-vertices. Implications of the ATLAS performance
 with respect to primary vertex reconstruction for the on-going and
 future physics analyses are discussed.
\end{Abstract}
\vfill
\begin{Presented}
2010 Hadron Collider Physics Symposium\\
Toronto, Canada,  August 23--27, 2010
\end{Presented}
\vfill
\end{titlepage}
\def\thefootnote{\fnsymbol{footnote}}
\setcounter{footnote}{0}
\section{ATLAS Inner Detector}
 The ATLAS Detector \cite{ATLASPerformanceBook} is a multi-purpose particle 
 detector in operation at the LHC at CERN. The aim of the ATLAS 
 experiment is to study $pp$ collisions at energies up 
 to $14$~TeV. The detector is composed of several sub-detectors
 designed to study a variety of physics processes. For the 
 reconstruction of vertices, the Inner Detector is of most 
 importance. The Inner Detector is made up of the high-resolution 
 semiconductor pixel and silicon microstrip (SCT) detectors and, 
 at higher radii with respect to the interaction point, of the straw-tube 
 tracking detectors (TRT). The entire setup is contained inside the superconducting 
 solenoidal magnet, producing a field of $2$~T.

 The Pixel detector consists of $3$ barrel layers, and $3$ end-cap disks either side of the 
 interaction point. The design resolution in $R\phi$ and $Z$ are of $10\;\mu{\rm m}$ and $115\;\mu{\rm m}$
 respectively.  The Semiconductor Tracker (SCT) consist of $4$ double-sided barrel layers, 
 and $9$ end-cap discs at each side of the interaction point, providing a resolution 
 of $17\;\mu{\rm m}$ and $580\;\mu{\rm m}$ in $R\phi$ and $Z$ respectively.
 The Transition Radiation Tracker is composed of $73$ layers of straw tube chambers in the barrel and
 $80$ layers in each end-cap providing $R\phi$ resolution of about $130\;\mu{\rm m}$.
 In summary, the Inner Detector provides around $3$ Pixel, $8$ SCT and $30$ TRT measurements per charged track.
 A detailed description of the Inner Detector and of its expected performance can be 
 found in \cite{ATLASPerformanceBook}. 
\section{Primary vertex reconstruction at $900$~GeV}
 During the first run of the LHC, the probability of $pp$ collisions to pile-up in the same bunch-crossing 
 was estimated to be $10^{-5}$. Consequently, the primary vertex finder was configured to
 reconstruct  exactly one primary vertex at each bunch-crossing. The tracks for the vertex reconstruction
 were pre-selected using the following criteria: measured transverse momentum $p_T > 150$~MeV;
 transverse impact parameter with respect to the beam position  $|d0| < 4$~mm;
 errors on measured transverse and longitudinal  impact parameters: $\sigma (d0) < 0.9$~mm and
 $\sigma (z0) < 10$~mm respectively; at least $4$ SCT hits; each track has at least $6$ hits in silicon
 detectors and at least $1$ Pixel hit.

 A dedicated Billoir-like \cite{Billoir} fitter was used to fit the vertices. While the beam-spot information
 was used for the pre-selection, the beam-spot constraint was not applied during the fit. The tracks
 least compatible with the vertex estimate are removed one by one, starting from the highest $\chi ^2$ contribution.
 The process is repeated until no tracks incompatible with the current vertex estimate is left.
\subsection{Analysis of charged particle multiplicities at $900$~GeV}
\begin{figure}[htb]
 \centering
\includegraphics[height=1.6in]{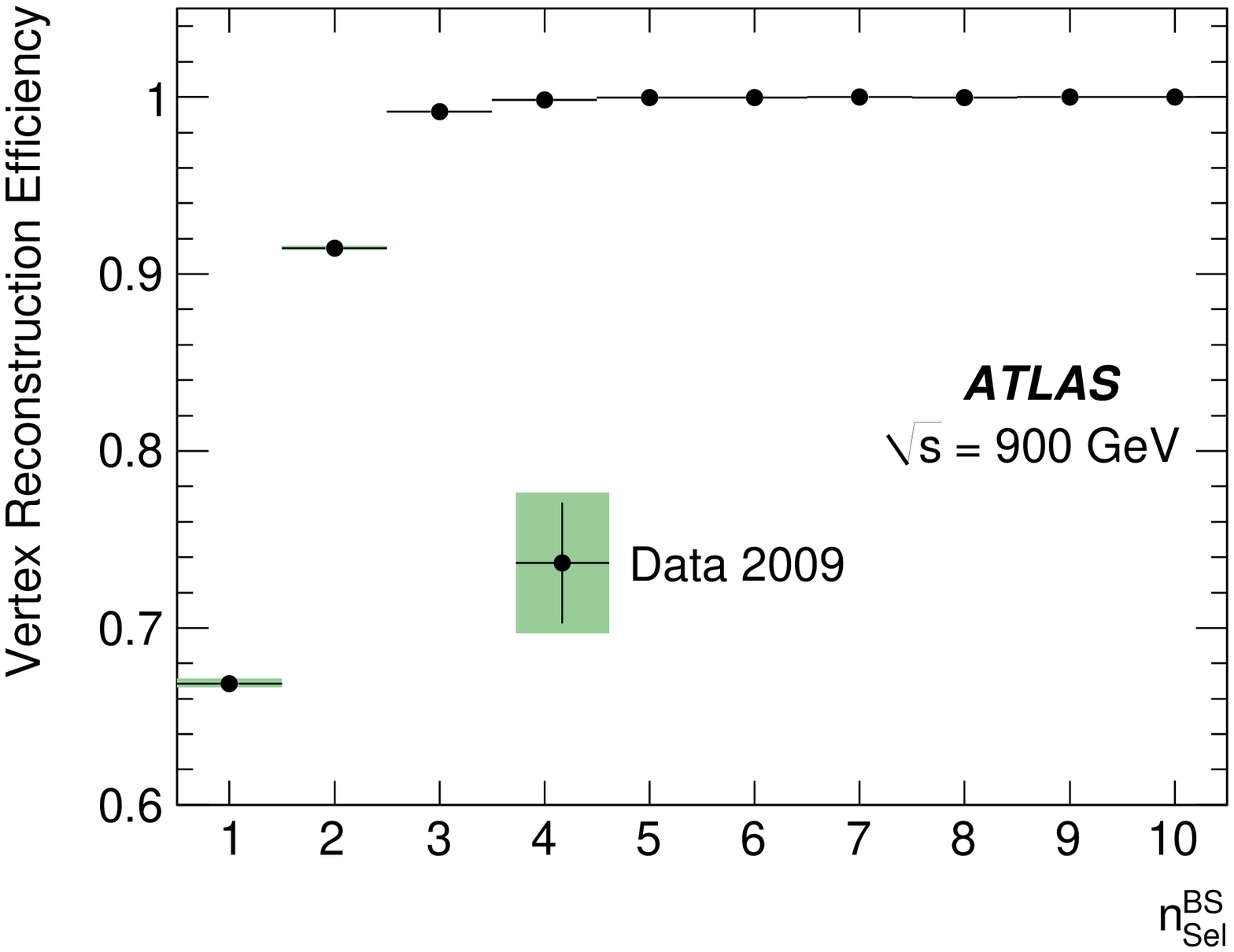}
\includegraphics[height=1.6in]{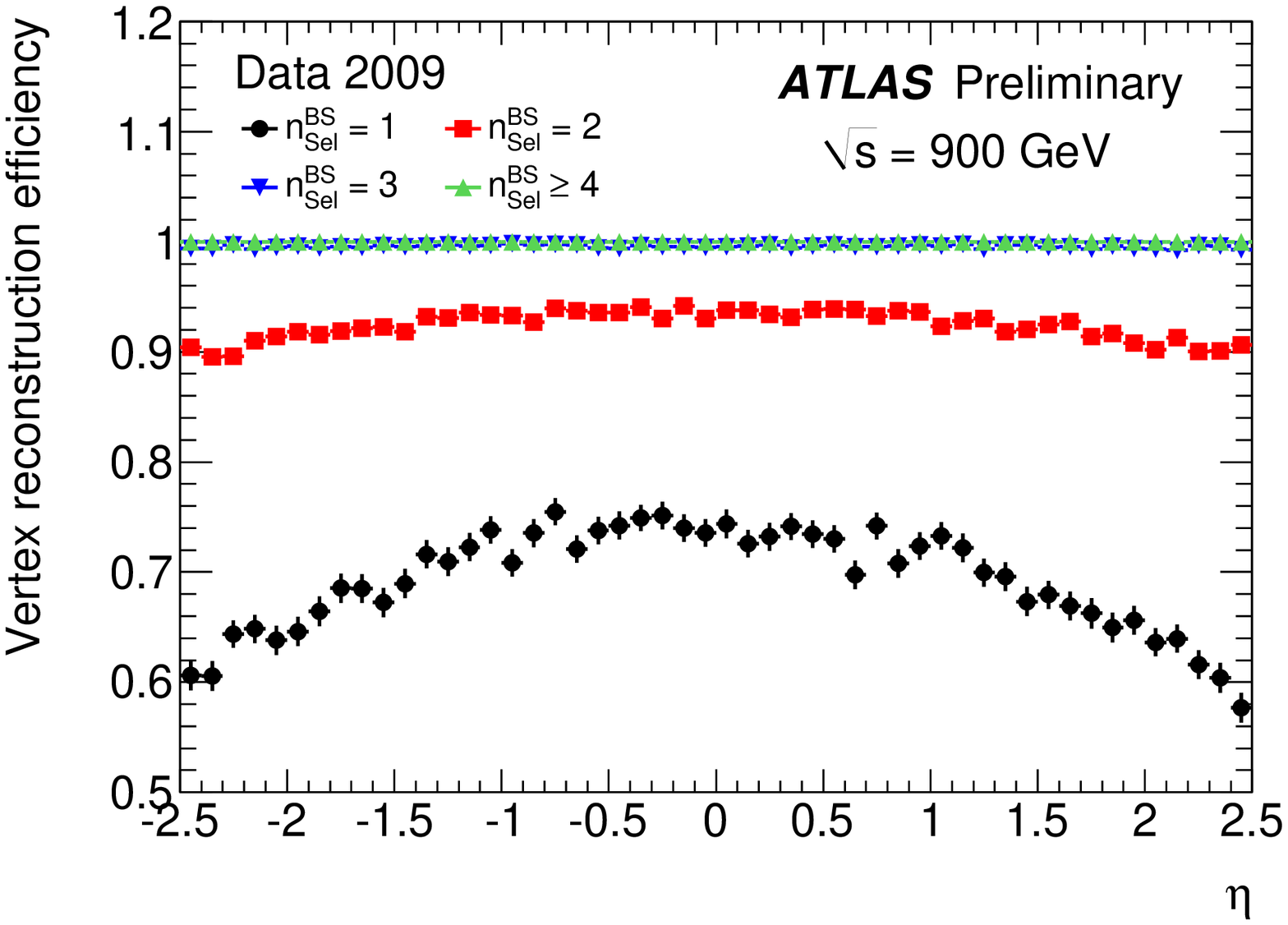}
 \caption{Efficiency of the vertex reconstruction as a function of the number of tracks
 selected for the analysis at $900$~GeV (left). Efficiency of the vertex reconstruction 
for one, two and three selected tracks as a function of the pseudorapidity $\eta$ (right).}
 \label{fig:eff900}
\end{figure}

 The measurement of the charged particle multiplicities at $900$~GeV was the first ATLAS analysis
 based on the collision data. The details of the analysis can be found in \cite{mb1}. The final
 measured distributions were corrected for the vertex reconstruction efficiency which
 was measured with respect to the Level-1 Minimum Bias Trigger Scintillators \cite{mb1}.
 The efficiency was measured as a function of the number of primary charged tracks selected for the
 final analysis. The impact parameter cuts with respect to the primary vertex were replaced by the
 cuts with respect to the beam-spot position. The only identified  source of systematic errors
 was found to  be due to the beam background.  Shown in Fig.~\ref{fig:eff900} (left) is
 the efficiency of the primary vertex reconstruction as a function of number of tracks
 selected for the analysis.
 It can be noted that the efficiency of $100\%$ is reached already with
 four selected tracks. The inefficiencies for fewer than four can be explained
 by the cuts imposed on the quality of the reconstructed vertices: only vertices
 with three and more fitted tracks  are accepted for the analysis.
 Shown in Fig.~\ref{fig:eff900} (right) is the efficiency of the vertex reconstruction
 for one, two and three selected tracks as a function of the pseudorapidity $\eta$.
 An $\eta$-dependency of the efficiency was observed for the case of one selected track and
 was corrected for in the analysis.
\section{Primary vertex reconstruction at $7$~TeV}
 For the first data collected at $7$~TeV, the amount of
 the pile-up was estimated as $10^{-3}$, which  was no longer negligible.  A new reconstruction
 strategy, based on the iterative vertex finder and adaptive vertex fitter \cite{Rudi} was used.
 The transverse momentum requirement for charged particle tracks was lowered to  $p_T> 100$~MeV.
 
 First, exactly one vertex is fitted from all the pre-selected tracks. Then, tracks
 incompatible by more than $7\sigma$ with this initial estimate are used to seed and reconstruct
 a new vertex candidate. This process is repeated until all available tracks are used or no
 new vertex seed can be created. The beam-spot information was used to constrain the vertex fit.
 The pile-up events were identified as triggered  bunch-crossings where at least one
 additional primary vertex with at least $4$ fitted tracks is reconstructed.
\subsection{Analysis of charged particle multiplicities at $7$~TeV}
\begin{figure}[htb]
 \centering
 \includegraphics[height=1.6in]{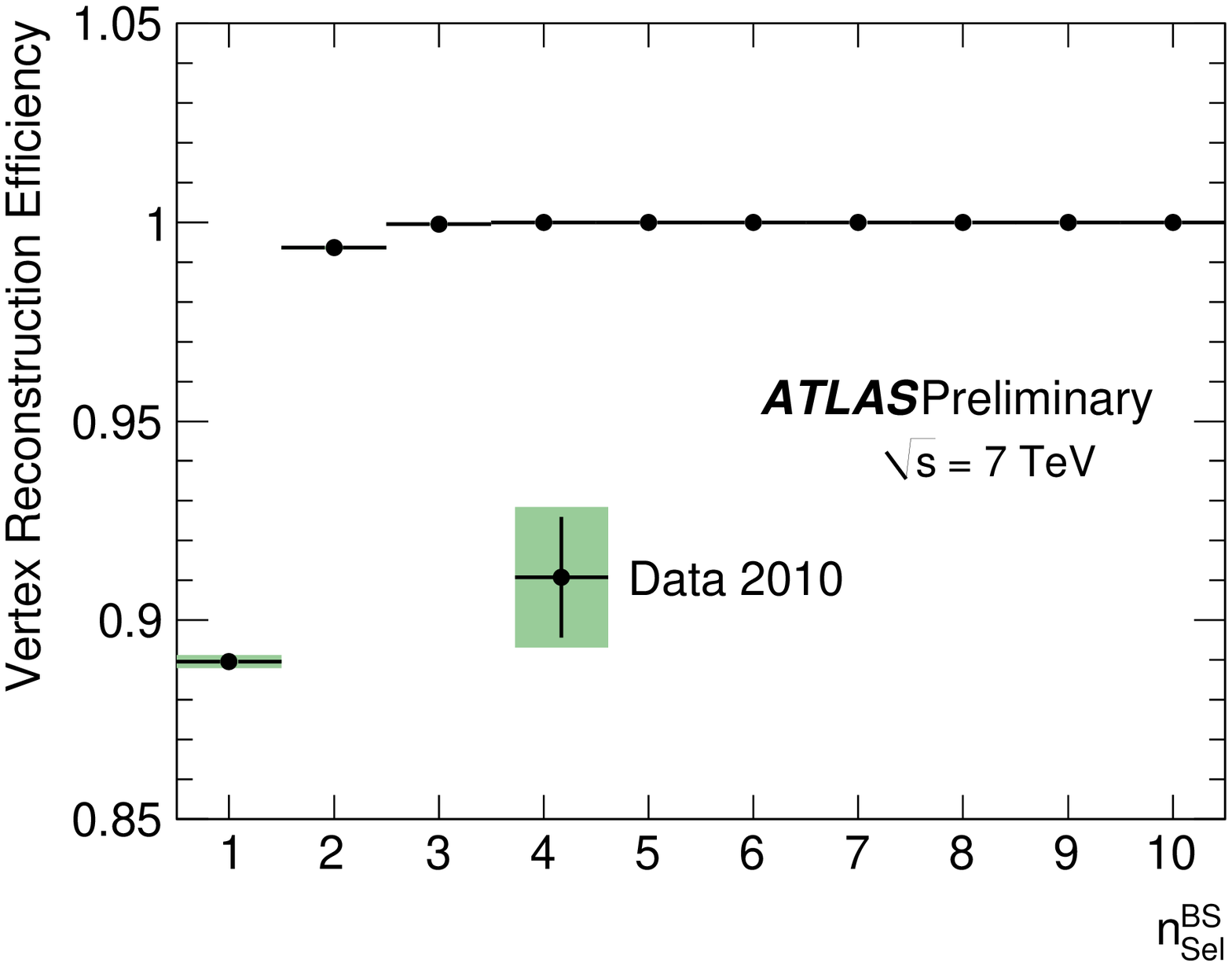}
 \includegraphics[height=1.6in]{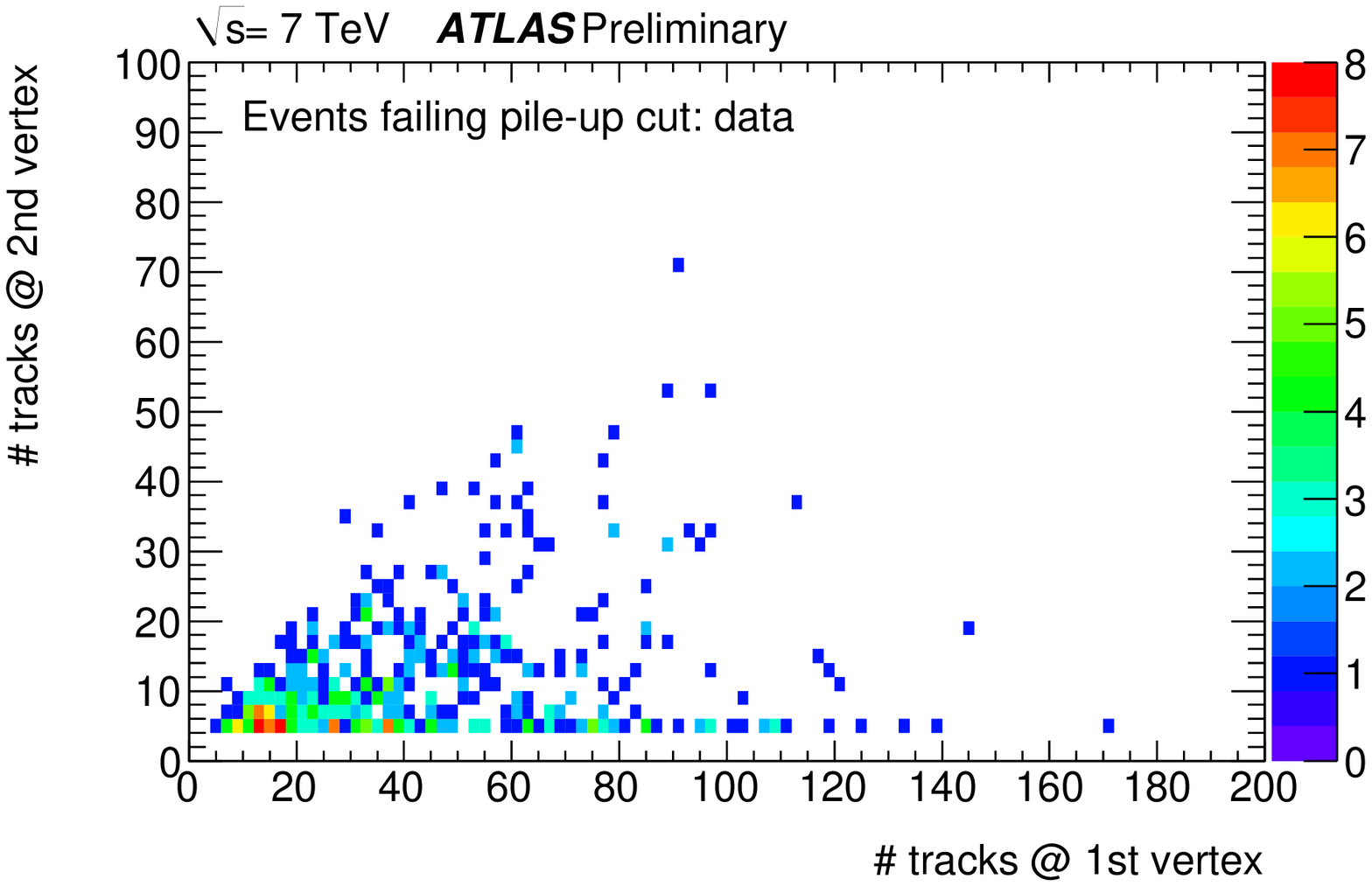}
 \caption{Efficiency of the vertex reconstruction as a function of the number of tracks
 selected for the analysis at $7$~TeV (left). Number of tracks fitted to 
the second vertex as a function of number of tracks fitted to the first vertex
for the events where more than one primary vertex was reconstructed (right).}
 \label{fig:eff7}
\end{figure}

 Shown  in Fig.~\ref{fig:eff7} (left) is the efficiency of the
 primary vertex reconstruction as a function of the number of tracks
 selected for the analysis of charged particle multiplicities at $7$~TeV \cite{mb15}.
 It can be noted that starting from three tracks, the efficiency reaches about $100$\%.
 Comparing to the analysis at $900$~GeV, smaller inefficiencies for low numbers of tracks
 are observed. This can be explained by higher track multiplicities in $7$~TeV
 collisions and  lower $p_T$ threshold selected for this analysis.
 In addition, only the  vertices with two and more fitted
 tracks were selected for this analysis.
 The efficiency was calculated only for the events where no pile-up 
 was identified. The corresponding systematic uncertainty was found to be tiny.

 Shown in Fig.~\ref{fig:eff7} (right) is the number of tracks fitted to
 the second vertex as a function of number of tracks fitted to the first vertex
 for the events where more than one primary vertex was reconstructed. It can be
 noted that in a fraction of the events, the second primary vertex is produced.
 This may happen because the primary vertex has been split by the reconstruction algorithm
 or because of the misidentificantion of secondary
 vertices. The majority of such events are removed from consideration by
 applying the $4$-track requirement mentioned above. The corresponding
 systematic effects are carefully evaluated and accounted for.
\section{Primary vertex resolutions at $7$~TeV}
 The resolutions on the positions of reconstructed primary vertices were
 calculated by  correcting the  reconstructed primary vertex errors by
 corresponding scale factors. Sets of tracks participating in the primary
 vertex reconstruction were randomly split in halves and corresponding pairs
 of vertices were reconstructed for each event. Gaussian fits were made to
 the distributions of distance pulls between split vertices. The scale 
 factors were defined as the standard  deviations of these fits.
 The measurement of the primary vertex position resolution is discussed in detail in \cite{res}.

\begin{figure}[htb]
 \centering
\includegraphics[height=1.6in]{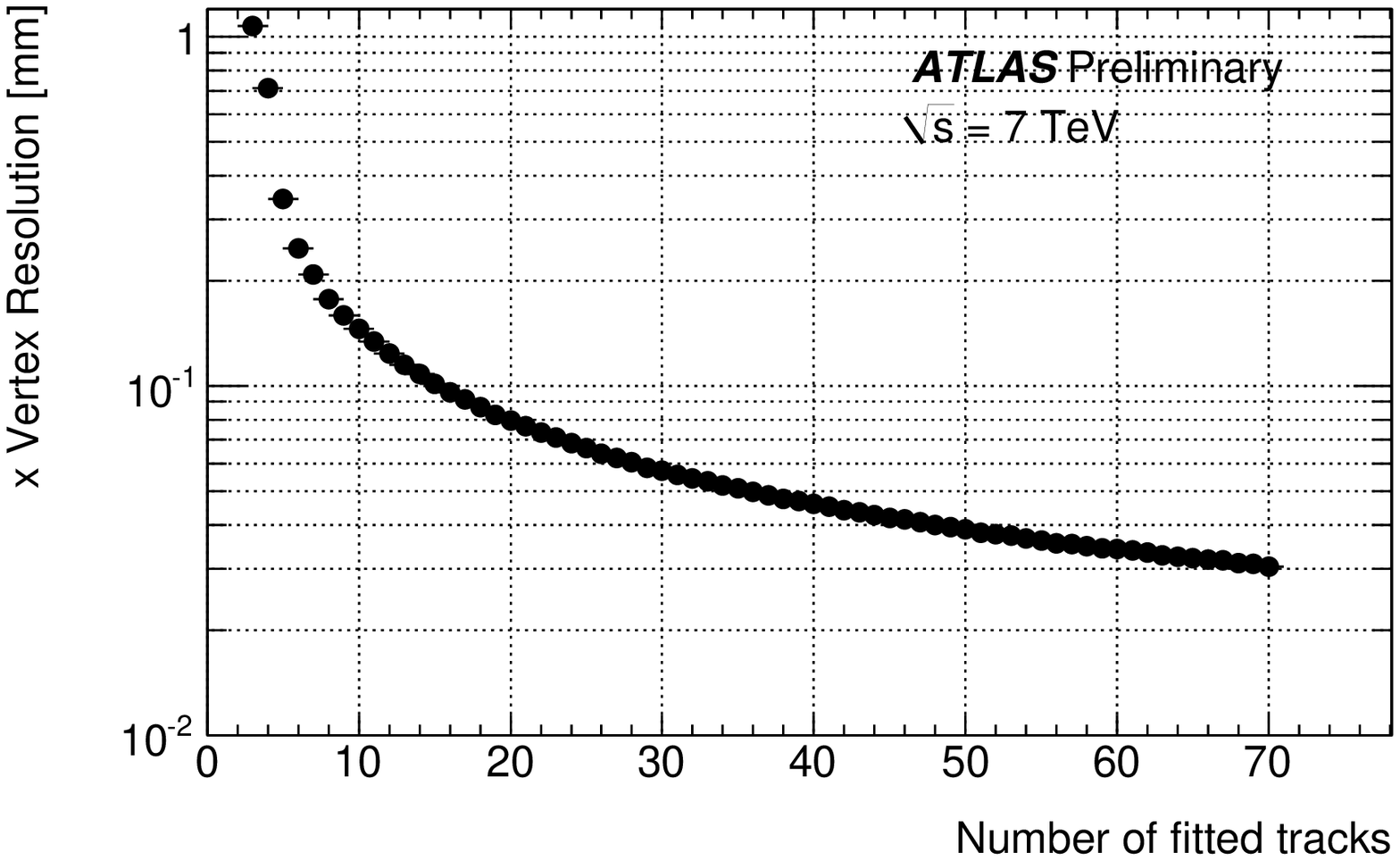}
\includegraphics[height=1.6in]{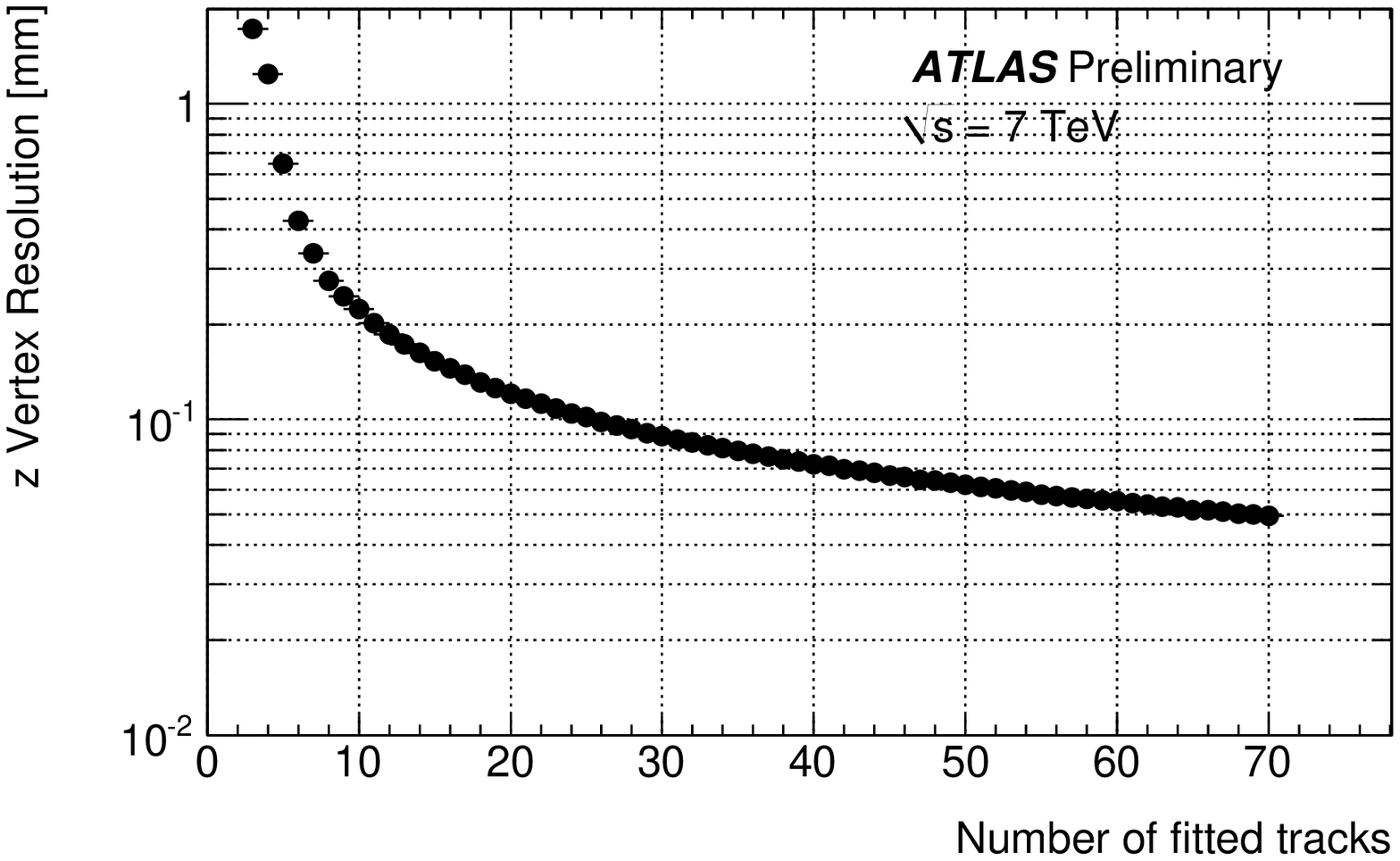}
 \caption{ Resolutions of the $X$ and $Z$
 coordinates of reconstructed primary vertices as a function of a total number of
 tracks fitted to a vertex.}
 \label{fig:res}
\end{figure}

 Shown in Fig.~\ref{fig:res} are the measured resolutions of the $X$ and $Z$
 coordinates of reconstructed primary vertices as a function of the total number of
 tracks fitted to a vertex. It can be observed that the position resolutions reach 
 the values of about $30\;\mu{\rm m}$ in $X$ and $50\;\mu{\rm m}$ in $Z$ for more 
 than $60$ fitted tracks. These values match the expectations for the $7$~TeV 
 minimum bias collisions with the $p_T$ threshold of $100$~MeV.

\section{Conclusions}
The ATLAS experiment at the LHC has started to take data in the Autumn of 2009. 
Different strategies for the reconstruction of primary vertices were applied 
during the $900$~GeV and $7$~TeV runs of the LHC. Reconstruction efficiencies 
close to $100$\% are achieved in both cases. For the $7$~TeV run, the coordinate
resolutions of primary vertices reach $30\;\mu{\rm m}$ and $50\;\mu{\rm m}$ in 
planes transverse and longitudinal to the beam axis respectively for high multiplicity 
minimum bias events.

\end{document}